\documentclass{appolb}
\pdfoutput=1

\usepackage{graphicx}
\usepackage{cite}

\usepackage{xcolor}
\definecolor{cBlue}{RGB}{0,110,191}
\usepackage[pdftitle={Lightcone expansion beyond leading power},hypertexnames=true]{hyperref}
\hypersetup{bookmarksnumbered,colorlinks,
    linkcolor={black},
    citecolor={cBlue},
    urlcolor={cBlue}}
\usepackage{slashed,mathtools}
\newcommand{\klammer}[1]{ \left( #1 \right)  }

\usepackage{stmaryrd}

\usepackage{amsmath,amssymb,amsfonts}

\def\slash#1{#1 \hskip-0.45em /}

\begin{document}
\eqsec 
\title{
Lightcone expansion beyond leading power%
\thanks{Presented at Matter to the Deepest 2023}
}
\author{Sebastian Jaskiewicz
\address{Albert Einstein Center for Fundamental Physics, Institut f\"ur Theoretische Physik, Universit\"at Bern, Sidlerstrasse 5, CH-3012 Bern, Switzerland}
 }
\maketitle
\begin{abstract}
\noindent We discuss recent developments in descriptions of processes using power expansion around the lightcone within Soft-Collinear Effective Theory. First, we present an overview  of the systematically improvable framework that enables factorization of high-energy scattering processes beyond leading power in the expansion in ratios of energy scales. As an illustration of the relevant concepts, we describe the recently derived factorization theorem for the off-diagonal channel of the Drell-Yan production process at threshold. This example exposes endpoint divergences appearing in convolution integrals in factorization formulas. Lastly, we discuss the solution to these complications   developed  in the context of ``gluon thrust'' in $e^+e^-$ collisions.

\end{abstract}
  
\section{Introduction}

Factorization theorems describing the decoupling of physical phenomena occurring at disparate energy scales and knowledge of universal  objects governing singular limits of scattering processes are integral in enabling accurate theoretical predictions in collider physics. Perhaps the best-known example is the factorization of sufficiently inclusive scattering cross-sections  in hadronic collisions into a perturbatively calculable short-distance part and parton distribution functions (PDFs), which capture the low-energy behaviour \cite{Collins:1985ue}. 
Despite the long history, most factorization theorems are formulated  only at the leading power (LP) in the expansion in ratios of the disparate energy scales. 
Focusing on efforts utilising effective field theory methods, in this contribution, we discuss the recent advancements made towards descriptions valid at subleading powers. 

Studies of subleading power corrections are important for phenomenological applications \cite{Beneke:2019mua,vanBeekveld:2021hhv}. As has been shown recently for the case of inclusive Higgs production via gluon fusion, the size of the leading logarithmic (LL) corrections at next-to-leading power (NLP) can rival those of next-to-next-to-leading logarithmic (NNLL) at leading power \cite{vanBeekveld:2021hhv}. Therefore, if towers of leading power logarithms  are included to high logarithmic accuracy, the subleading power logarithms should also be included at least at the leading logarithmic order to maintain control over the genuine size of the errors in the predictions. 

On top of the phenomenological applications, 
investigations of power corrections present also an intriguing challenge from the theoretical perspective and plenty of work has recently been carried out on this topic in various contexts. A non-exhaustive list of recent studies in literature includes investigations of Higgs production in gluon fusion at threshold, deep-inelastic  scattering (DIS) at large Bjorken-$x$,  hadronic $e^+e^-$ annihilation, and the  threshold Drell-Yan (DY) process
\cite{Bonocore:2016awd,Moult:2018jjd,vanBeekveld:2021hhv,Beneke:2018gvs,Beneke:2019oqx,Moult:2019mog,Bahjat-Abbas:2019fqa,Beneke:2019mua,Broggio:2023pbu,Moult:2019uhz,Moult:2019vou,Ajjath:2020ulr,Beneke:2020ibj,Broggio:2021fnr,Beneke:2022obx,vanBijleveld:2023vck,vanBeekveld:2021mxn,Inglis-Whalen:2021bea,AH:2020xll}.
Furhtermore, studies have been carried out for the 
single Higgs boson production and decay amplitudes \cite{Penin:2014msa,Liu:2017axv,Liu:2017vkm,Liu:2018czl,Wang:2019mym,Liu:2019oav,Liu:2020eqe,Anastasiou:2020vkr,Liu:2021chn,Liu:2022ajh,Liu:2020tzd,Liu:2020wbn,Bodwin:2021cpx,Bodwin:2021epw}. 
Advancements beyond leading power have also been achieved 
for variables such as N-jettiness \cite{Boughezal:2016zws,Moult:2016fqy,Moult:2017jsg,Ebert:2018lzn,Boughezal:2018mvf,Boughezal:2019ggi,Vita:2024ypr}, the $q_T$ of the Higgs boson or the lepton pair \cite{Ebert:2018gsn,Cieri:2019tfv,Oleari:2020wvt,Buonocore:2023mne,Ferrera:2023vsw}, 
and in the context of QED \cite{Bell:2022ott,Laenen:2020nrt,Engel:2021ccn,Engel:2023ifn,Engel:2023rxp,Balsach:2023ema}, and $B$ physics \cite{Beneke:2017vpq,Beneke:2019slt,Feldmann:2022ixt,Cornella:2022ubo,Hurth:2023paz}. 

We begin by giving a brief overview of the general subleading power considerations within soft-collinear effective theory (SCET) \cite{Bauer:2000ew,Bauer:2000yr,Bauer:2001yt,Beneke:2002ph,Beneke:2002ni}. Namely, we discuss the basis of subleading power operators, the expanded Lagrangian with power suppressed soft-collinear interactions, and process specific kinematic power corrections. 
In order to illustrate the relevant concepts in a concrete example, we will focus on the recent derivation of next-to-leading power  factorization formula for the quark-gluon channel of the Drell-Yan production process at threshold presented in \cite{Broggio:2023pbu}.  The obtained results expose the presence of endpoint divergences, which is a ubiquitous complication appearing   in  various subleading power factorization theorems. In the last part of this contribution, we will discuss a solution of this problem in the context of ``gluon thrust''  in $e^+e^-$ collisions using refactorization \cite{Beneke:2022obx}.

\section{Sources of power corrections}
\label{sec:formalism}
In the following discussion, we adopt the 
subleading power SCET 
formalism developed
in \cite{Beneke:2004in,Beneke:2017ztn,Beneke:2018rbh,Beneke:2019kgv,Beneke:2017mmf,Boer:2023yde}\footnote{
For alternative approach  of
constructing power suppressed operator 
basis in the label formulation of SCET see~\cite{Marcantonini:2008qn,Kolodrubetz:2016uim,Feige:2017zci,Moult:2017rpl}.}.
SCET describes the dynamics of soft   and collinear partons. 
The collinear partons contain a large momentum
component along one of the light-like directions, 
and are suppressed  
along the remaining ones. 
Therefore, it is convenient to use 
light-like reference vectors $n^{\mu}_{i-}$
and $n^{\mu}_{i+}$ for each of the collinear 
directions~$i$: 
	$n^{\mu}_{i-} = (1,\vec{n}_i),
	n^{\mu}_{i+} = (1,-\vec{n}_i)$. 
The $\vec{n}_i$ is a three-vector, and the light-like
 reference vectors  satisfy $n_{i-}\cdot n_{i+} =2$ and
  $n_{i-}^2=n_{i+}^2=0$.   We work in  ${{\text{SCET}}_{{\text{I}}}}$, where there is a strict hierarchy between the hard, (anti)collinear, and soft virtualities.

\begin{figure}[t]
 	\begin{center}
 		\includegraphics[width=7.5cm]{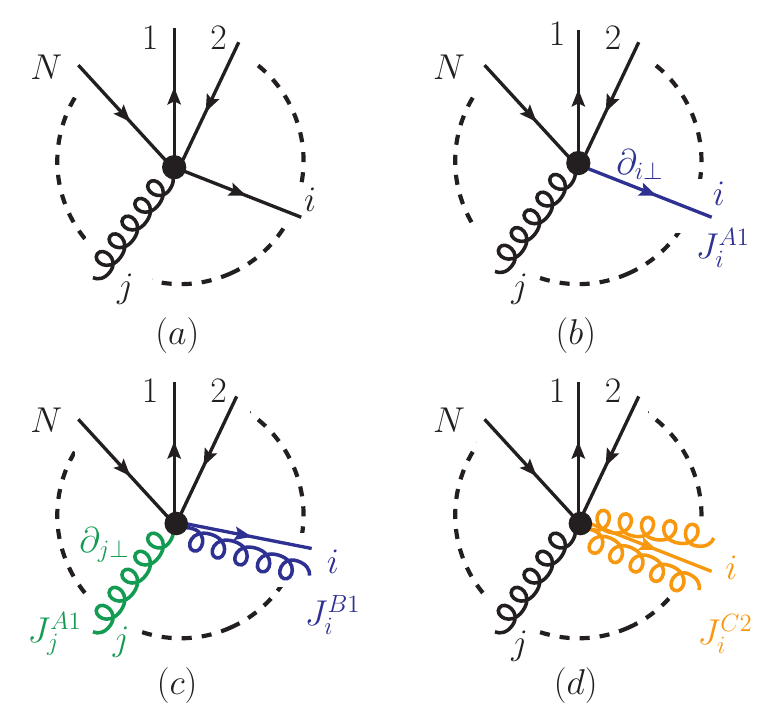}
 	\end{center}
 	\caption{	\label{fig:nlp_currents} Examples of possible contributions to the $N$-jet scattering process. Panel $(a)$ shows a leading power $N$-jet operator with one building block present in each of the $N$ collinear directions. 
 	Panels $(b)$, $(c)$, and $(d)$ depict possible power corrections to the leading power process in $(a)$, as described in the text. }
  \end{figure}

In the first step, hard modes are integrated out through a procedure which matches QCD to a basis of SCET operators. The SCET operators are constructed out of
collinear-gauge invariant
building blocks 
 \cite{Hill:2002vw} 
\begin{equation}
\label{gaugeInvBuildBlocks}
{\psi_{i}}(x)  \in
\quad
 \left\{ \begin{array}{ll}
\displaystyle 
\chi_i(x)=W_i^{\dagger}(x)\xi_i(x)  & {\quad{\text{$i$-collinear quark}}} \\[0.3cm]
\displaystyle 
\mathcal{A}^{\mu}_{i\perp}(x)= 
W_i^{\dagger}(x)\big[iD^{\mu}_{i\perp}W_i(x)\big]
 & {\quad{\text{$i$-collinear gluon}}} 
\end{array}\right.
\end{equation}
where $	\xi_i(x) = \frac{\slash{n}_{i-}\slash{n}_{i+}}{4}\psi_i(x)$,
$i D_i^{\mu}(x)= i\partial^{\mu} +g_sA_i^{\mu}(x)$,
and  the $i$-collinear Wilson line is defined as
\begin{equation}\label{collinearWilsonline}
W_i(x)= \textbf{P}\exp\bigg[ig_s\int_{-\infty}^{0} ds\,
 n_{i+}A_i(x+sn_{i+})   \bigg]\,.
\end{equation}
The symbol $\textbf{P}$ in the above equation denotes a path ordering operator. 
A generic $N$-jet operator has the following structure \cite{Beneke:2017ztn}  
\begin{eqnarray}\label{eq:NjetOperator}
	J= \int \bigg[\prod_{ik} dt_{i_k}\bigg]
	C(\{t_{i_k}\})\,J_s(0)
	 \prod_{i=1}^NJ_i(t_{i_1},t_{i_2}...) \,,
\end{eqnarray}  
where $C(\{t_{i_k}\})$ is a generalised Wilson coefficient which captures the hard modes,  $J_s$ is a soft operator, and $J_i$ is a product of $n_i$  collinear
building blocks associated with a specific collinear direction $n_{i+}^{\mu}$
\begin{eqnarray}
	J_i(t_{i_1},t_{i_2}...)=
\prod_{k=1}^{n_{i}}\psi_{i_k}(t_{i_k} n_{i+}).
\end{eqnarray}
Each of the collinear building blocks in \eqref{gaugeInvBuildBlocks}
has a scaling of $\mathcal{O}(\lambda)$ 
\cite{Beneke:2002ph}, where $\lambda$ is the small power counting parameter of the theory  and its specific form is dictated by the process under consideration. In the leading power configuration there is only a single building block present in each of the collinear directions, as depicted in an $N$-jet example in panel $(a)$ of figure~\ref{fig:nlp_currents}. 
There are two ways to extend the basis of operators to subleading powers \cite{Beneke:2017ztn}.  
The first is through the introduction of $\partial_{\perp}^{\mu}$ derivatives which act on the building blocks already present in the leading power configuration, bringing an $\mathcal{O}(\lambda)$ suppression. For example, $J^{A1}_i(t_i) =i\partial^{\mu}_{i\perp}\chi_i(t_in_{i+})$ as depicted in direction $i$ in panel $(b)$ of figure~\ref{fig:nlp_currents}. Secondly, additional building blocks can be added within each collinear direction. Since the building blocks scale as $\lambda$, each single insertion also induces an $\mathcal{O}(\lambda)$ suppression with respect to the leading power operator. Examples of this are shown in panels $(c)$ and $(d)$ of figure~\ref{fig:nlp_currents} where
$ J^{B1}_i(t_{i_1},t_{i_2})= \mathcal{A}^{\mu}_{i\perp}(t_{i_1}n_{i+})
  \chi_i(t_{i_2}n_{i+})$ and $J^{C2}_i(t_{i_1},t_{i_2},t_{i_3})=\mathcal{A}^{\mu}_{i\perp}(t_{i_1}n_{i+})\mathcal{A}^{\mu}_{i\perp}(t_{i_2}n_{i+})
  \chi_i(t_{i_3}n_{i+})$. 
The basis is organised into currents $J^{An}$, $J^{Bn}$, $ J^{Cn},\ldots$ where the letter $A,B,C,\ldots$ denotes the number of fields present in particular collinear direction, and the number $n$ gives the overall power suppression with respect to the leading power operator within each sector. From $\mathcal{O}(\lambda^3)$, purely soft currents such as soft quark field $q(x)$ appear in the basis. These are collected in $J_s$. Naturally, the 
sum of the power suppression from the different collinear sectors and the soft $J_s$ gives  the overall power suppression for the $N$-jet operator.  
  
Next, we consider the SCET Lagrangian for QCD which is separated into $N$ collinear parts and a global soft  term   
\begin{eqnarray}\label{scetL}
	\mathcal{L}^{}_{{\text{SCET}}_{{\text{I}}}}=
	\mathcal{L}^{}_{s}+
	\sum_{i=1}^{N}\mathcal{L}^{}_{i}\,.
\end{eqnarray}
The Lagrangian terms in the collinear sectors are each systematically  expanded in a small power counting parameter   $\lambda$. In general, we have
\begin{eqnarray}\label{powerexp}
 \mathcal{L}^{}_{i} = \underbrace{\mathcal{L}^{(0)}_{i}}_{\mathcal{O}(\lambda^0)}
 + \underbrace{\mathcal{L}^{(1)}_{i}}_{\mathcal{O}(\lambda^1)}
 +\underbrace{\mathcal{L}^{(2)}_{i}}_{\mathcal{O}(\lambda^2)}+\dots
\end{eqnarray}
The first term in the expansion, $\mathcal{L}^{(0)}_{i}$, is the leading power contribution, and the remaining terms are the power corrections. The specific form of the Lagrangian expanded to $\mathcal{O}(\lambda^2)$ is given in \cite{Beneke:2002ni}. Following the framework introduced in \cite{Beneke:2017ztn}, we adopt the interaction picture such that 
all operator matrix elements are evaluated with the leading-power 
SCET Hamiltonian, 
and the subleading power Lagrangian terms enter the basis as perturbations through time-ordered product insertions with lower power currents \cite{Beneke:2017ztn}.  As an example, at $\mathcal{O}(\lambda^{n+m})$ the time-ordered product
operators take the form 
\begin{eqnarray}\label{eq:TordProdDef}
	J^{T(n+m)}_i(t_i) = i \int d^dx \,T\left\{ 
J^{An}_i(t_i)\,\mathcal{L}^{(m)}_i(x)	
\right\}. 
\end{eqnarray}
Lastly, the power suppression for a process can be captured by the so-called kinematic correction. This type of correction is process specific and it originates in the phase-space approximations which are valid only up to leading power. The kinematic correction is made up of appropriately expanding to  the phase-space to subleading power accuracy.
See \cite{Beneke:2018gvs} for an example in the case of Drell-Yan production at threshold in the diagonal channel.

\section{Threshold factorization of the Drell-Yan quark-gluon channel}
Using the general construction outlined in the previous section, we now 
focus on the Drell-Yan process at threshold  and
set the number of collinear directions $N=2$. 
The threshold limit 
is characterised by $z \equiv Q^2/\hat s\to 1$, where 
$Q^2$ is the invariant mass of the 
final-state lepton pair, and $\hat s$ is the 
partonic centre-of-mass energy squared.
Specifically, we consider the off-diagonal 
 processes 
$g\bar{q}\,(qg) \to \gamma^* + X$. The derivation of the 
subleading power factorization theorem for this process 
has recently been presented in \cite{Broggio:2023pbu,Jaskiewicz:2021cfw}.
In this contribution we discuss important considerations and the key features. For details, we refer the interested reader to 
\cite{Broggio:2023pbu}
where all the  necessary ingredients to validate the NLP factorization to next-to-next-to-leading order have been computed. 

We consider the invariant 
mass distribution  for the  Drell-Yan process given by 
\begin{equation}
\frac{d\sigma_{\rm DY}}{dQ^2} = \sigma_0 \sum_{a,b} \int_{\tau}^1\, \frac{dz}{z}\, 
{\cal L}_{ab}\bigg(\frac{\tau}{z}\bigg)\, \Delta_{ab}(z) + \mathcal{O}\left(\frac{\Lambda}{Q}\right), \qquad  \sigma_0 = \frac{4\pi \alpha_{\rm em}^2}{3 N_c Q^2 s},
\label{eq:dsigsqDelta}
\end{equation}
where ${\cal L}_{ab}(y)$ is the standard parton luminosity function, $\tau = Q^2/s$, and $\Lambda$ is the confinement scale of QCD. We seek to obtain the factorization theorem
for the partonic part $\Delta_{g\bar q}(z)$.  

\begin{figure}[t]
\begin{center}
\includegraphics[width=0.42\textwidth]{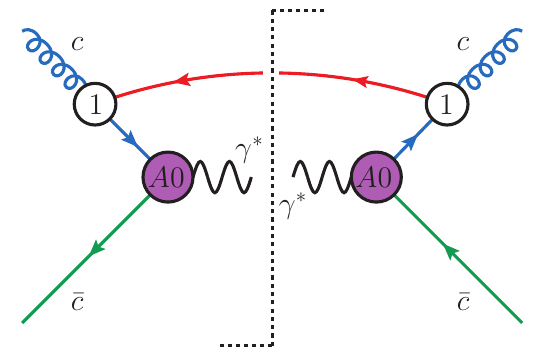} 
\end{center}
\caption{Lowest-power 
contribution to the gluon-antiquark channel in the DY process. 
The incoming collinear gluon is converted 
into a collinear quark via an emission of a 
soft anti-quark. In the effective field theory, this 
interaction is captured by a time-ordered product 
insertion of the power suppressed Lagrangian term ${\cal L}_{\xi q}^{(1)}$ 
with the LP SCET current $J^{A0}$. 
This power suppressed interaction vertex is denoted
by the index ``1'' in the picture. $A0$ is the LP hard matching coefficient.}
\label{fig:XSnondecv}
\end{figure}

In the threshold limit, the final state is forced to contain only the soft radiation and the hard matching to SCET operators can be carried out at amplitude level.
In this limit, for the process 
$g\bar{q}\,\rightarrow\,\gamma^*+X$ to 
occur, the 
incoming gluon drawn from the parton distribution function must be converted 
to a threshold collinear quark via the emission of a soft antiquark.
The interaction between collinear fields and soft quarks is inherently a 
subleading power effect appearing for the first time in the SCET Lagrangian at $\mathcal{O}(\lambda)$ in 
$\mathcal{L}^{(1)}_{\xi q}$
\begin{equation} 
\mathcal{L}^{(1)}_{\xi q} = 
\bar{q}^{\,+}\slash{\mathcal{A}}^{(0)}_{c\perp}\chi_c^{(0)}+ {\rm h.c.},
\end{equation}
where we have already performed the decoupling transformation  \cite{Bauer:2001yt}, and $\chi^{(0)}_c(z) = Y_+^{\dagger}(z_-) \chi_c(z)$,
$\mathcal{A}^{(0) \mu}_c(z) =
Y_+^\dagger(z_-)\mathcal{A}^{\mu}_c(z)Y_+(z_-)$, ${q}^{\pm} = Y_{\pm}^{\dagger}\, q_s$ with the soft Wilson lines defined in the following way 
\begin{eqnarray}
Y_{\pm}\left(x\right)&=&\mathbf{P}
\exp\left[ig_s\int_{-\infty}^{0}ds\,n_{\mp}
A_{s}\left(x+sn_{\mp}\right)\right].
\end{eqnarray}
Hence, in contrast to the 
$q\bar{q}$-channel presented in 
\cite{Beneke:2019oqx}, the contribution 
to the DY cross-section from the 
$g\bar{q}$-channel appears for the first time at NLP. It occurs via a 
time-ordered product insertion  of 
$\mathcal{L}^{(1)}_{\xi q}$ with the LP current, as shown in 
Fig. \ref{fig:XSnondecv}.
The derivation of the factorization theorem proceeds as follows. The matching of the  electromagnetic quark current to SCET fields is carried out at LP using 
 \begin{eqnarray}\label{eq:LPmatching}
\bar{\psi} \gamma_\rho \psi(0) = \int dt\, 
d\bar{t}\,\widetilde{C}^{A0,A0}(t,\bar{t}\,) \,
J_\rho^{A0,A0}(t,\bar{t}\,) \,,
 \end{eqnarray}
where, {\emph{after}} decoupling, in the notation for $N$-jet operators described above
\begin{eqnarray}\label{eq:LPcurrent}
J_\rho^{A0,A0}(t,\bar{t}\,) =  \bar{\chi}_{\bar{c}} 
(\bar{t} n_-)Y^{\dagger}_{-}(0)\gamma_{\perp\rho} Y_{+}(0)\chi_c(t n_+ )\,.
 \end{eqnarray}
We suppressed the superscript $(0)$ on decoupled fields in the equation above.
Power suppression is then generated via the aforementioned 
time-ordered product insertions of ${\cal L}_{\xi q}^{(1)}$. 
These insertions give rise to a new feature in subleading 
power factorization theorems with respect to their leading 
power counterparts. Namely, the additional collinear fields
introduced via time-ordered product insertions are too
 energetic to be accepted in the final state in the threshold
 kinematics. Therefore, an amplitude level matching to PDF-collinear
 fields must first be performed which gives rise to \emph{NLP collinear functions}  \cite{Beneke:2018gvs,Beneke:2019oqx,Moult:2019mog}. For the case of 
 $g\bar{q}$-channel of DY, the matching is done onto the PDF-collinear
 gluon which gives rise to the following collinear matching equation \cite{Jaskiewicz:2021cfw,Broggio:2023pbu}
\begin{eqnarray}\label{m1qg}
i \int d^dz \,{T}\Big[ 
\chi_{c,\gamma f}\left(tn_+\right)
\,\mathcal{L}^{(1)}_{\xi q}(z) \Big]
= \frac{2\pi}{g_s}  \int \frac{d\omega}{2\pi}
\int \frac{dn_+p}{2\pi} \,e^{-i(n_+p)t} 
\int \frac{dn_+p_a}{2\pi} && \nonumber \\
&& \hspace{-12cm}
\times \, {G}^{\eta,A}_{\xi q;\gamma\alpha,fa}
\left(n_+p,n_+p_a; \omega \right)\, 
\hat{\mathcal{A}}^{{\rm PDF}\,A}_{c \perp \eta}(n_+p_a)
\int dz_-\, e^{-i\omega\,z_-}
\,\mathfrak{s}_{\xi q;\alpha,a}(z_-).
\end{eqnarray}
In the above equation, $\eta$ is a Lorentz  index, $\alpha$ and $\gamma$ are  Dirac  indices, and  $a,f$ and $A$ are a fundamental and adjoint  colour indices respectively.
The soft structure, $\mathfrak{s}_{\xi q}$,  originates from $\mathcal{L}^{(1)}_{\xi q}$.
Making the indices explicit,  it reads 
\begin{eqnarray}\label{qgsoftstructure}
\mathfrak{s}_{\xi q;\alpha,a}(z_-) = \frac{g_s}{in_-\partial_z}q^{+}_{\alpha,a}(z_-)\,.
\end{eqnarray}
After this step, the derivation of the factorization formula proceeds in the standard manner. The amplitude is squared, and the sum over the PDF-(anti)collinear state is carried out. The matrix element of the PDF-(anti) collinear fields is expressed in terms of the standard parton distributions. 
The gluon-antiquark contribution to DY begins at NLP, so the phase-space can be truncated at leading power, yielding no kinematic corrections in this channel.
The final NLP 
factorization formula can be simplified using generic properties of the collinear 
function. Namely, we can write
\begin{equation}
\label{collFunction4}
{G}_{\xi q,\gamma\alpha,fa}^{\eta,A}(n_+p,  n_+p_a;\omega) = 
{G}_{\xi q}(n_+p;\omega)\,\delta\big(n_+p-n_+p_a\big)\, 
\mathbf{T}^{A}_{fa}\,\left[\slashed{n}_{-} \gamma^{\eta}_{\perp}\right]_{\gamma\alpha}\,. 
\end{equation}
After some further manipulations, we arrive at the final result \cite{Broggio:2023pbu}
\begin{equation}
    \label{eq:QGnlpfact}
\Delta_{g\bar{q}}|_{\rm NLP}(z) = 
8 H(Q^2)\int {d\omega}\, d\omega' \,  
G^*_{\xi q}(x_an_+p_A;\omega') \,
G_{\xi q}(x_an_+p_A;\omega) 
\, S(\Omega,\omega, \omega') \,.
\end{equation}
In this equation, $H(Q^2)$, is the well-known LP hard function,  given by the 
square of the LP short-distance 
coefficient: $H(\hat{s})
=|C^{A0,A0}(x_an_+p_A,x_bn_-p_B)|^2 $ $
= H(Q^2)+\mathcal{O}(\lambda^2)$. The 
soft function ${S}(\Omega,\omega,\omega')$ 
is given by
\begin{eqnarray}\label{softoperator2}
S(\Omega,\omega, \omega') &=&  
\int \frac{dx^0}{4\pi} \int \frac{dz_{-}}{2\pi} 
\int \frac{dz'_{-}}{2\pi}e^{-i\omega z_{-}}
e^{+i\omega' z'_{-}}e^{+i \,\Omega \, x^0/2} 
\nonumber\\ &&\times \frac{1}{N_c^2-1}
\langle 0| \bar T \Big(
\frac{g_s}{in_-\partial_{z'}}
\bar{q}^{\,+}(x^0+z'_{-})\,\mathbf{ T}^{A}\,
 \big\{Y^{\dagger}_{+}(x^0)Y_{-}(x^0)\big\}\Big)
\nonumber\\&& \times  \frac{\slashed{n}_{-}}{4} 
T\Big(\left\{Y^{\dagger}_{-}(0)Y_{+}(0)\right\} 
\, \mathbf{ T}^{A}\,\frac{g_s}{in_-\partial_{z}} 
q^{+}(z_{-}) \Big) |0\rangle \,.
\end{eqnarray}
\begin{figure}[t]
\begin{center}
\includegraphics[width=0.75\textwidth]{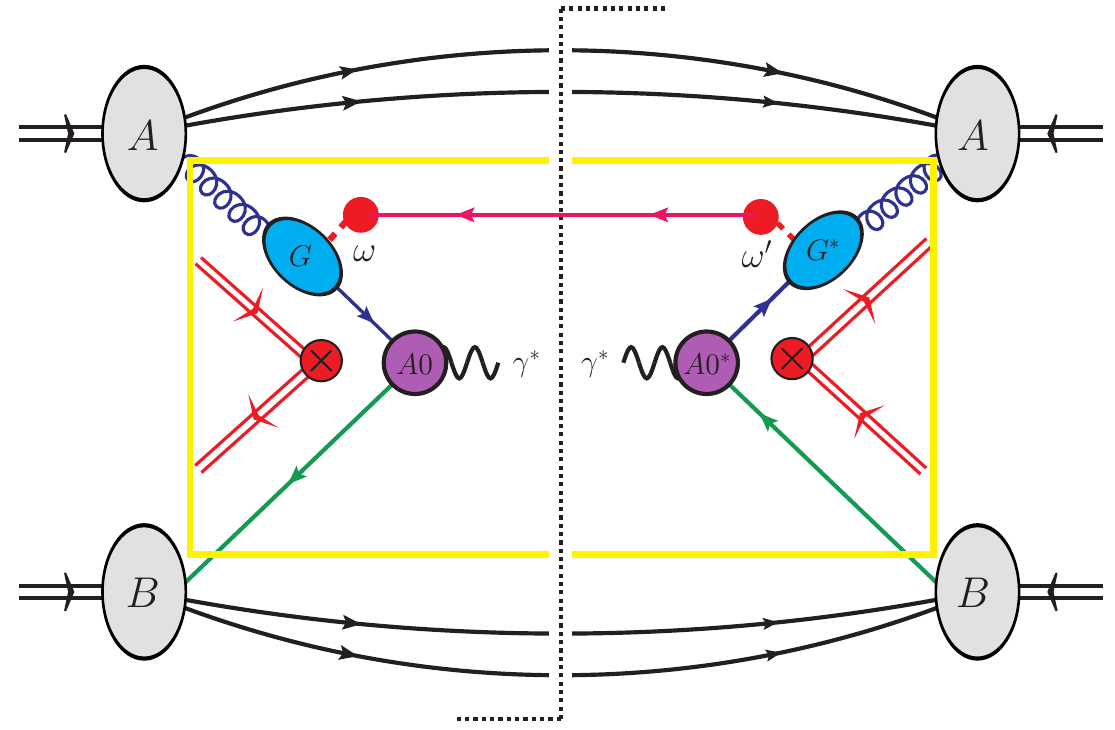} 
\end{center}
\caption{Schematic representation of the 
factorization theorem in \eqref{eq:QGnlpfact}.  
The red lines represent soft fields, blue lines 
represent collinear fields, and green lines 
are the anti-collinear fields. 
The leading power 
short-distance coefficient $C^{A0}$ and its complex conjugate 
are the purple circles 
labelled with ``$A0$'' and ``$A0^*$'',  respectively.  
NLP collinear functions are represented by the  blue ovals denoted by $G$, $G^*$.
Lastly, $\omega$ and $\omega'$ are the 
convolution variables between the 
soft and collinear 
functions.}
\label{fig:XSfactorized}
\end{figure} 
This concludes our overview of the derivation of the 
factorization theorem for the off-diagonal channel contribution to DY at NLP. Keeping the dimensional regulator in place, this result has been validated up to NNLO in \cite{Broggio:2023pbu} via an explicit computation 
of each of the ingredients appearing in the above result to the required perturbative accuracy:  soft functions to $\mathcal{O}(\alpha_s^2)$, collinear functions at $\mathcal{O}(\alpha_s)$.\footnote{Equivalent collinear functions have been calculated to $\mathcal{O}(\alpha_s^2)$ in the context of $H \to gg$ amplitude in \cite{Liu:2021mac}.} The hard function is known up to three-loops \cite{Gehrmann:2010ue}. 
Using the obtained results, it is also possible to expose the endpoint divergences appearing in this example. Namely, focusing on the collinear and soft piece we have 
\begin{eqnarray}
 \int_0^\Omega d\omega \,\underbrace{\big(n_+p
 	\, \omega\big)^{-\epsilon}}_{{\rm{collinear\,piece}}}
 \,\underbrace{\frac{1}{\omega^{1+\epsilon}}
 	\frac{1}{(\Omega-\omega)^{\epsilon}}}_{{\rm{soft\,piece}}}
 \,.
\end{eqnarray} 
It is apparent that the integral is well defined when exact $\epsilon$ dependence is retained in the integrand. However, if the soft and collinear pieces are first expanded individually in the context of resummation, it is clear that a  divergent integral is
encountered. Namely, in this case, we find a term proportional to $\int d\omega \delta (\omega) \ln(\omega)$. Therefore, the standard renormalization procedure and four-dimensional convolutions do not in general yield the correct structure of the NLP logarithms of $(1-z)$. In the $g\bar{q}$-channel investigated here the issue arises already at the leading logarithmic level, whereas it appears for the first time at next-to-leading logarithmic order in diagonal channels \cite{Beneke:2019oqx}. As mentioned in the introduction, the appearance of endpoint divergences in subleading power factorization theorems is a ubiquitous issue studied in a variety of contexts \cite{Moult:2019uhz,Beneke:2020ibj,Liu:2019oav,Liu:2020tzd,Liu:2020wbn,Liu:2022ajh,Bell:2022ott,Feldmann:2022ixt,Cornella:2022ubo,Hurth:2023paz,delCastillo:2023rng,Beneke:2022obx}. In the next section, we discuss a solution for this problem using refactorization ideas developed for ``gluon thrust''  \cite{Beneke:2022obx}.

\section{Refactorization in ``gluon thrust'' }
In this section, we switch focus to $e^+e^-$ collisions and consider the ``gluon thrust'' event shape where at leading order the gluon recoils a quark-antiquark pair
\begin{equation}\label{eq:gluonthrust}
e^+ e^-\to\gamma^*\to [g]_{c} + [q\bar q]_{\bar c}\,.
\end{equation}
The thrust variable $T$ is defined as  \cite{Brandt:1964sa,PhysRevLett.39.1587}
\begin{align}
T=\mbox{max}_{\vec n}\,\frac{\sum_{i}
\left|\vec{p_{i}}\cdot\vec{n}\right|}
{\sum_{i}\left|\vec{p_{i}}\right|},
\end{align}
where the index $i$ runs over hadrons (partons) in the final state.
In the limit $\tau=1-T \to 0$,\, 
back-to-back jets are formed by the partons, and 
large logarithms in the $ \tau$ variable appear at all orders in $\alpha_s$. 
The quark-antiquark two-jet process
contributing at LP in the $\tau$ expansion  is known to high logarithmic accuracy~\cite{Becher:2008cf}. Comparatively, much less is understood about the process in eq.~\eqref{eq:gluonthrust} beginning at NLP in the $\tau$ expansion,
where the leading term is $\alpha_s\ln \tau$.

At the first order in $\alpha_s$, 
the gluon jet can be induced in two ways:  
\begin{enumerate}
\item[I] The large anti-collinear 
momentum can be carried away by both the quark and the anti-quark,  which create a single jet that recoils against the collinear gluon.
\item[II] The collinear gluon momentum is balanced by either the anti-collinear quark or anti-quark, which renders the remaining fermion soft.
\end{enumerate}
\begin{figure}
\begin{center}
\includegraphics[width=0.3\textwidth]{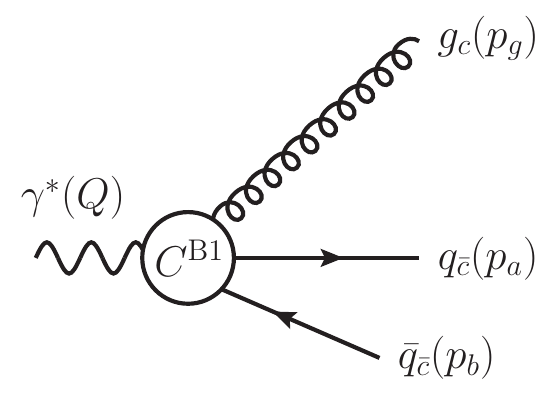}\hspace{0.15cm}
\includegraphics[width=0.3\textwidth]{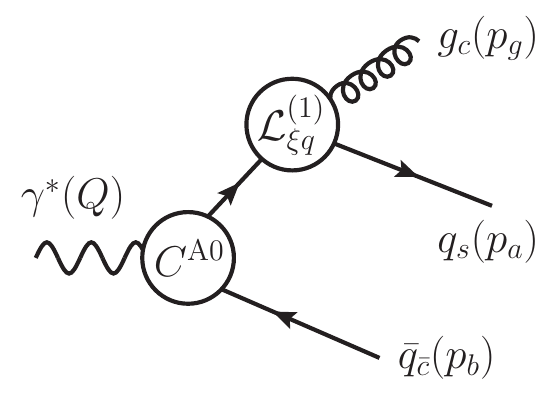}\hspace{0.15cm}
\includegraphics[width=0.3\textwidth]{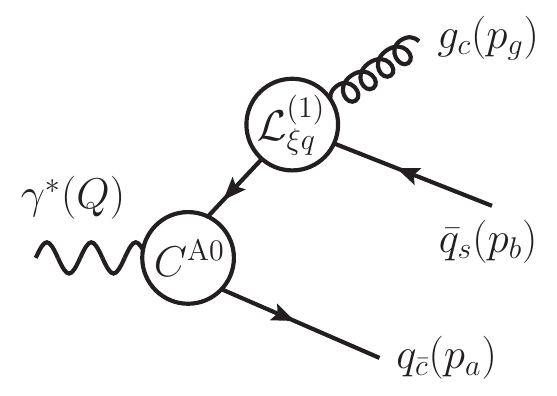}
\caption{\label{fig:SCETfig} SCET representation of the 
``gluon thrust'' amplitude in the two-jet limit. 
}
\end{center}
\end{figure}
From the SCET point of view, in situation 
I the $q\bar{q} g$ state is directly produced by hard scattering operator as described in section~\ref{sec:formalism}. Due to the subleading operator involved, we refer to this situation   as the ``B-type'' scenario and it is  depicted in the left-most diagram of figure~\ref{fig:SCETfig}. In this case, momentum
conservation fixes 
only the total momentum of the $q\bar q$ pair. The amplitude in turn depends on the fractions of the anti-collinear momentum carried by the individual particles. 
If one of these fractions becomes  
small, the corresponding parton becomes soft and must instead be counted
as contributing to possibility II, also referred to as the ``A-type'' scenario. In this contribution,  the hard scattering vertex is the LP
current and the full momentum of the
quark (or anti-quark) is then transferred to the gluon, which renders the
daughter fermion soft. Following the discussion in section~\ref{sec:formalism}, this situation is captured by a time-ordered product insertion of
$\mathcal{L}^{(1)}_{\xi q}$ 
which describes soft (anti)quark emission. The ``A-type'' scenario is shown in the middle and right-most diagrams of figure~\ref{fig:SCETfig}.

We begin with the factorization theorem for two-hemisphere invariant mass
distribution of ``gluon thrust'' in Laplace space: 
\begin{align}
\frac{1}{\sigma_{0}}\frac{\widetilde{d\sigma}}{ds_R ds_L}
&=\int  d\omega d\omega'\,
\left|C^{A0}\right |^{2}\times
\mathcal{J}^{(\bar{q})}_{\bar{c}}\times 
\mathcal{J}_{c}\left(\omega,\omega'\right)
\otimes S_{\rm NLP} \left(\omega,\omega'\right)
\nonumber \\& \hspace{-1cm}+
\int  drdr'\,
C^{B1}(r) C^{B1}(r')^{*}\otimes
\mathcal{J}^{q\bar q}_{\bar{c}}\left(r,r'\right)\times 
\mathcal{J}_{c}^{(g)}\times S^{(g)}\,.
\label{eq:ffschematic}
\end{align}
$C$ denotes the hard matching coefficients, $\mathcal{J}$ the
jet functions,  and $S$ are the soft functions.
Only the dependence on convolution variables
which give rise to divergent integrals is retained. 
The $r$, $r'$
convolution integrals diverge logarithmically  for
$r,r'\to 0,1$, and the $\omega$,$\omega'$  convolutions
are logarithmically divergent for $\omega,\omega'\to \infty$.  

The soft quark from scenario II is contained in $S_{\rm NLP}$. 
In the overlap region,   
soft momentum $\omega$ carried by the soft quark is actually {\em large} and could count as
belonging to $\mathcal{J}^{q\bar q}_{\bar{c}}$, with a {\em small} 
fraction~$r$ of the anti-collinear momentum.
Removing this quark from 
$S_{\rm NLP}$ reduces it to $S^{(g)}$. Hence, taking all these changes into account, the 
hard process effectively changes from $A0$-type to $B1$-type. 
In these limits,
 we can  identify 
$r=\omega/Q$, $r'=\omega'/Q$, such that
the  {\em integrands} of the two terms in \eqref{eq:ffschematic} become identical.  
We can then perform a rearrangement at the level of the {\em integrand}, such that both terms are separately finite.
From this point,  
standard RG techniques can be utilized to resum the logarithms appearing
in the soft, jet, and hard  functions.   Details of this procedure are presented in \cite{Beneke:2022obx}, which we summarize briefly below. 

\begin{figure}
\begin{center}
\includegraphics[width=0.4\textwidth]{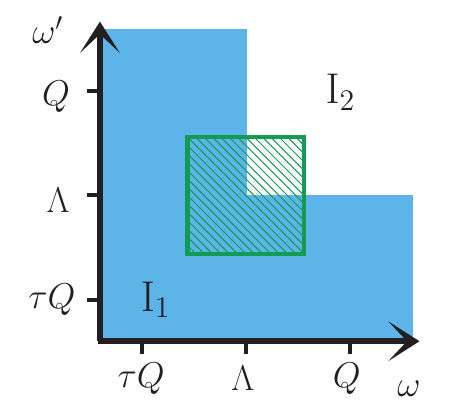}
\includegraphics[width=0.52\textwidth]{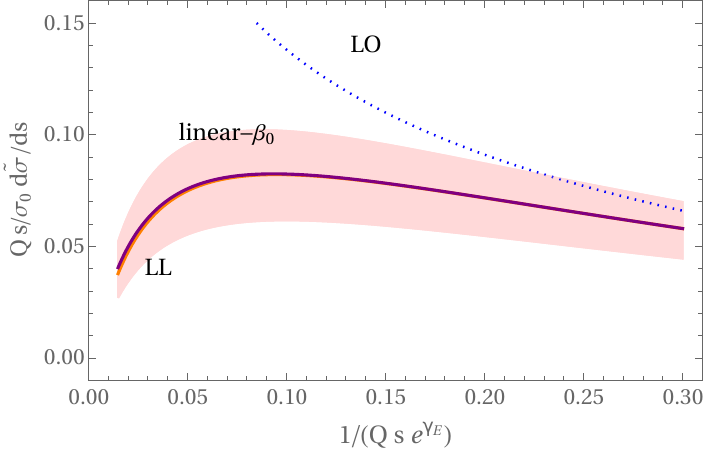}
\caption{\label{fig:overlap}
The left panel shows the split of 
\eqref{eq:scaleless_integral} into $I_1+I_2$  in the 
$\omega-\omega'$ plane as described in the text  
below eq. \eqref{eq:scaleless_integral}.  The right panel, displays the 
Laplace-space  gluon thrust distribution at LL. The light-red band is created by  the variation of the  initial scales as described in the text.  
}
\end{center}
\end{figure}

The rearrangement of endpoint-singular terms is achieved by
introduction of the following scaleless integral
\begin{eqnarray}
\hspace{-0.2cm}
\frac{2 C_F}{Q} \,f(\epsilon)\,|C^{\rm A0}|^2 \widetilde{\mathcal{J}}^{(\bar q)}_{\bar{c}}  
\widetilde{\mathcal{J}}_{c}^{(g)}  
\int_0^\infty d\omega d\omega'\,
\frac{D^{\rm B1} }{\omega}
\frac{D^{\rm B1*} }{\omega'} 
\left \llbracket \widetilde{S}_{\rm NLP}(s_R,s_L,\omega,\omega')\right 
\rrbracket,
\label{eq:scaleless_integral}
\end{eqnarray}
where we have suppressed the arguments in the $D^{B1}$ functions defined in 
eq. (57) of \cite{Beneke:2022obx}.
This integral is separated into two terms $I_{1,2}$ where $I_1+I_2=0$. The term $I_1$  is defined by $\omega$ {\em or} $\omega'$ smaller than a parameter $\Lambda$ and the complement region is denoted as $I_2$.  This split is depicted on the left-hand side of figure~\ref{fig:overlap}. 

The endpoint rearrangement 
entails taking away the $I_1$ piece from the B-type contribution 
and  the $I_2$ term from the A-type contribution. The resulting 
expressions are individually endpoint-finite. The
dependence on $\Lambda$ cancels exactly between the 
two terms if no further approximations are made.
The RG equations for the objects appearing in the 
factorization formulas can now be solved as presented in sections 5.1 and
5.2 of \cite{Beneke:2022obx}. The end result 
for the LL accurate resummed expression in Laplace space is:
\begin{eqnarray}
	\label{eq:mainLLresult}
	\frac{1}{\sigma_0}\frac{\widetilde{ d\sigma}}{ ds_R ds_L}|_{\rm LL}&=&
	2\cdot \frac{2 C_F}{Q s_R}\frac{\alpha_s(\mu_c)}{4\pi}\,
	\,\exp\left[4C_FS\klammer{\mu_h,\mu_{\bar{c}}}
	+4C_AS\klammer{\mu_s,\mu_c}\right]
	\nonumber\\
	&&\hspace*{-3cm}
	\times	\klammer{\frac{1}{s_Ls_R e^{2\gamma_E}\mu_s^2}}^{\!-2C_A A\klammer{\mu_s,\mu_c}}
	 \int_\sigma^Q\frac{d\omega}{\omega}\,
	\klammer{\frac{\omega }{s_R e^{\gamma_E} \mu_{s\Lambda}^2}}^{\!-2\klammer{C_F-C_A}A\klammer{\mu_{s\Lambda},\mu_{h\Lambda}}}
	\\[0.2cm]\nonumber
	&&\hspace*{-3.0cm}\times
 \klammer{\frac{Q^2}{\mu_h^2}}^{\!-2C_F A\klammer{\mu_h,\mu_{\bar{c}}}}
		e^{\left[4\klammer{C_F-C_A}S\klammer{\mu_{s\Lambda},\mu_{h\Lambda}}\right]}
	\,\klammer{s_R e^{\gamma_E} Q}^{2C_FA\klammer{\mu_{h\Lambda},\mu_{\bar{c}}}+2C_A A\klammer{\mu_c,\mu_{h\Lambda}}}.
\end{eqnarray}
The functions $S\klammer{\nu,\mu}$ and $A_{\gamma_i}\klammer{\nu,\mu}$ are defined  in \cite{Neubert:2004dd}. To study the importance of  
{\em next}-to-leading logarithms, we vary  the
matching scales around the values adopted in \eqref{eq:mainLLresult}. Three pairs of scales are varied $(\mu_h, \mu_{h\Lambda})$, 
$(\mu_c,\mu_{\bar c})$, $(\mu_s, \mu_{s\Lambda})$ by a factor of $1/2$ 
and 2. Scale variation is computed by taking the resulting minimum and maximum values.  
Effect of this procedure is shown for  
$\frac{Q s}{\sigma_0}\frac{\widetilde{ d\sigma}}{ ds}$ in the 
right-hand panel of figure~\ref{fig:overlap} as the light-red 
band around the red curve (LL) which represents 
\eqref{eq:mainLLresult}. We also display the leading order and 
linear-$\beta_0$ truncation
of the leading logarithmic expression for comparison. 
The resulting sizeable scale variation emphasizes the necessity of  
resummation at NLL order. The endpoint-rearranged factorization formula derived in \cite{Beneke:2022obx} is the 
starting point for this systematically improvable analysis.

\section{Summary}
In this contribution, we highlighted the recent progress made towards achieving descriptions of factorization of physical scattering processes valid beyond leading power in the expansion in ratios of energy scales. In section~\ref{sec:formalism}, we presented an overview of the SCET formalism  
which  enables systematic inclusion of subleading power corrections in these processes. In section~3, we discussed the relevant concepts based on a concrete example of the factorization formula recently derived for the off-diagonal channel of the DY  process at threshold. Using the relevant results, we discussed the  appearance of 
endpoint divergences, and finally,   we reviewed the recently developed solution in the context of ``gluon thrust'' in $e^+e^-$ collisions. As is evident, studies of subleading power corrections present intriguing challenges from the theoretical perspective and  are phenomenologically relevant for the upcoming precision era of the LHC and the Electron  Ion Collider.

\bibliographystyle{JHEP}

\providecommand{\href}[2]{#2}\begingroup\raggedright\endgroup

\end{document}